\documentclass{aa-multiplecitationsrepair}

\usepackage{graphicx}
\usepackage{txfonts}
\usepackage{makecell}
\usepackage{bm}
\usepackage{placeins}

\usepackage{color,xcolor,xspace,pdflscape,dblfloatfix}

\definecolor{mhi}{rgb}{0.6,0.0,0.6}

\newcommand{\fig}[1]{Fig.~\ref{fig:#1}}

\newcommand{\tab}[1]{Table~\ref{tab:#1}}

\newcommand{\sect}[1]{Sect.~\ref{sec:#1}}
\newcommand{\app}[1]{Appendix~\ref{app:#1}}

\newcommand{\lcdm}[0]{$\Lambda$CDM\xspace}
\DeclareMathOperator{\sech}{sech}

\begin{document}

\title{Fornax dwarf spheroidal in MOND: its formation and the survival of its globular clusters}

\titlerunning{Globular clusters of Fornax in MOND}
\authorrunning{Michal B\'ilek \& Hongsheng Zhao}

   \author{
   Michal B\'ilek \inst{1,2},
   Hongsheng Zhao \inst{2,3,4,5}
   }

   \institute{Astronomical Observatory of Belgrade, Volgina 7, 11060 Belgrade, Serbia,     \email{michal.bilek@aob.rs}
    \and Scottish Universities Physics Alliance, University of Saint Andrews, North Haugh, Saint Andrews, Fife, KY16 9SS, United Kingdom,     \email{hz4@st-andrews.ac.uk}
    \and Imperial Centre for Inference and Cosmology (ICIC), Department of Physics, Imperial College, Blackett Laboratory, Prince Consort Road, London SW7 2AZ, U.K.
    \and CAS Key Laboratory for Research in Galaxies and Cosmology, Department of Astronomy, University of Science and Technology of China, Hefei 230026, China 
    \and Department of Astronomy, University of Science and Technology of China, Hefei 230026, People's Republic of China
             }
   \date{Received: January 2024; accepted March 2024}

 \abstract{
The Fornax dwarf spheroidal galaxy has five massive globular clusters (GCs). They are often used for testing different dark matter and modified gravity theories, because it is difficult to reconcile their old stellar ages with the short time they need to settle in the center of the galaxy due to dynamical friction. Using high resolution $N$-body simulations with the Phantom of Ramses code, we investigate whether the GCs of Fornax can be reconciled with the modified Newtonian dynamics (MOND), namely its QUMOND formulation. Observational data interpreted in MOND indicate that  Fornax is a tidal dwarf galaxy formed at redshift $z=0.9$ in a  flyby of the Milky Way (MW) and Andromeda galaxies, and that its GCs were initially massive star clusters in the disk of the MW. This helps us to set up and interpret the simulations. In the simulations, a point-mass GC orbits Fornax, and they both orbit the MW. When we ran multiple simulations with varying initial conditions for the GC, we found a 20\% probability of Fornax being observed with five unsunk GCs. The unsunk GCs have the observed radial distribution. Moreover, we found: 1) In MOND, Fornax has an orbit around the MW such that the pericenters coincide with the observed peaks in the star formation history of Fornax; 2) The simulations reproduce the observed ``diffuse stellar halo'' of Fornax; 3) The simulations predict that Fornax has a stellar stream, which could be detectable in the existing data. 4) An extra simulation shows that if Fornax was initially a rotating disky tidal dwarf galaxy, the gravitational influence of the MW would be  able to transform it into a nonrotating spheroidal. 5) Sometimes Phantom of Ramses does not conserve angular momentum. This makes the GC sink too fast if it is simulated as an $N$-body object.
 }

               
   \maketitle
\section{Introduction}
The Fornax dwarf spheroidal galaxy (Fornax hereafter), a satellite of the Milky Way (MW), hosts five ``classical'' globular clusters (GCs) and one ``new'', whose existence has been confirmed relatively recently  \citep{pace21}. The new GC has a much younger stellar age and lower mass than the classical ones.  The classical GCs, which we will discuss hereafter (without the word ``classical''), provide an interesting constraint on the essence of the dark sector: it turns out to be difficult to reconcile their high stellar ages, about 10\,Gyr, with the estimates of the time necessary for their settling in the center of Fornax due to dynamical friction \citep{tremaine76,hernandez98,oh00}. Dynamical friction is a process that transfers the orbital energy and angular momentum of the GCs into the internal energy and angular momentum of Fornax and the GCs.  

In the context of Newtonian gravity, various solutions to the mystery of the GCs of Fornax have been proposed. For example, it has been  proposed that the dark matter halo of Fornax has a core \citep{goerdt06},  the GCs were formed much further from Fornax than currently observed \citep{angus09}, Fornax experienced a major merger \citep{leung20}, or that the GCs are embedded in dark matter minihalos \citep{boldrini20}. A comprehensive model of galaxy formation in the \lcdm cosmology even shows that a few percent of the simulated dwarf spheroidals indeed possess GC systems similar to that of Fornax without any special adjustments \citep{shao21}. However, it should be pointed out that regardless of the advances in the problem of the GCs of Fornax, the \lcdm cosmology struggles to explain the existence of the disk of satellites of the MW, of which Fornax is clearly a member (see \sect{tdg} for details).

In this contribution, we investigate the survivability of the GCs of Fornax in modified Newtonian dynamics  (MOND, \citealp{milg83a}), namely in its QUMOND formulation \citep{qumond}. Initial analytic calculations indicated that the problem of survival of the GCs of Fornax (and other dwarf galaxies) is even fiercer in MOND than with Newtonian gravity and dark matter, because of more efficient dynamical friction \citep{ciotti04,sanchezsalcedo06,nipoti08}. It has been argued, again from analytic calculations, that the problem can be solved if GCs start their existence near the tidal radius of Fornax \citep{angus09}. Using high-resolution simulations of GCs in ultra-diffuse
galaxies, \citet{bil21} demonstrate that the analytic formulas are correct, but only as long as the GC is not too close to the center of the galaxy. Then the sinking is slowed down by the core-stalling effect \citep{hernandez98,read06,petts15}, which was ignored when deriving the analytical formulas for dynamical friction and is difficult to incorporate into them. That work came to the conclusion that MOND can explain the presence of GCs in ultra-diffuse galaxies. The simulations of \citet{bil24} investigate the survival of GCs in isolated gas-rich galaxies (90\% of baryons in gas)  both for MOND and \lcdm.  It turned out that, in MOND, supernova (SN) explosions cause large fluctuations of gravitational potential, which prevent the GCs from sinking, but only if the mass of the GC is less than $4\times 10^5\,M_\odot$. It remained unclear, however, to what extent this limit is  influenced by the numerical implementation of the baryonic processes. There were hints that the rate of SN explosions was smaller than it should.

{The main goal of the present work is to investigate the compatibility of QUMOND with Fornax GCs. To that end, we use high-resolution $N$-body simulations of point-mass GCs in a spherical Fornax model  without hydrodynamics. Compared to the case of the isolated ultra-diffuse galaxies in \citet{bil24}, Fornax has a smaller size, lower mass, and is gravitationally perturbed by the MW. It turns out that, for a proper setup on the simulation and its  interpretation, a wider context of the formation of Fornax and its GCs has to be taken into account. That is deduced mostly from observations and partly from older simulations and dynamical models. Apart from answering our main question, the simulations also give many other interesting results, which we also describe here. } 

\begin{table}
\caption{Parameters of Fornax, its GCs, the MW, and the observing frame.}
\label{tab:obs}
\centering
\begin{tabular}{lll}
\hline\hline                   
Parameter  & Value  & Note  \\
\hline 
& & \\
 \multicolumn{3}{c}{Fornax}  \\
\hline \hline 
Heliocentric distance & 143\,kpc & (1) \\
Angular scale & 24\arcmin/kpc, 2.5\,kpc/\degr & \\
Stellar mass & $3\times10^7\,M_\sun$ & (2) \\
Right ascension & 2h39m50.9s & (3) \\
Declination & -34$\degr$30$\arcmin$54$\arcsec$ & (3)\\
S\'ersic index & 0.8 & (3) \\
Effective radius & 0.75\,kpc & (3) \\
Heliocentric proper motion $\mu_{\alpha cos(\delta)}$ & 0.381\,mas\,yr$^{-1}$ & (4)\\
Heliocentric proper motion $\mu_\delta$ & -0.358\,mas\,yr$^{-1}$ & (4)\\
Heliocentric radial velocity & 55.3 km\,s$^{-1}$ & (5) \\
\hline 
& & \\
 \multicolumn{3}{c}{GCs of Fornax}  \\
\hline \hline  
Names & \hspace{-7em} Fornax(1, 2, 3, 4, 5) & \\
Stellar masses & \hspace{-7em} (0.4, 1.5, 5.0, 0.8, 1.9) $\times 10^5\,M_\sun$ & (6) \\ 
Projected distances & \hspace{-7em}  (1.67, 0.93, 0.64, 0.15, 1.65)\,kpc & (7) \\
\hline
& & \\
 \multicolumn{3}{c}{Milky Way}  \\
\hline \hline  
Stellar mass & \hspace{-4em} $6\times10^{10}\,M_\sun$ & (8)\\
\hline
& & \\
 \multicolumn{3}{c}{Observing frame}  \\
\hline \hline  
Galactic center position & \hspace{-4em} [8.27, 0, 0]\,kpc & (9) \\
Galactic center velocity &  \hspace{-4em} [-9.3, -251.5, -8.59]\,km\,s$^{-1}$ & (9) \\
\hline   
\end{tabular}
\tablefoot{(1) \citet{oakes22}. (2) Intermediate value between \citet{mcconnachie12} (stellar mass obtained from the $V$-band luminosity by assuming the mass-to-light ratio of one) and \citet{deboer12} (from  a stellar populations synthesis model). (3) Sample S0  of \citet{yang22}, effective radius converted from the provided S\'ersic radius. (4) \citet{battaglia22}. (5) \citet{mcconnachie12}. (6) \citet{deboer16}. 
(7)  Calculated from the coordinates of the GCs by \citet{mackey03} and the coordinates and distance of Fornax given above. The distances differ from those given by  \citet{mackey03} because a different center of Fornax is used. (8) \citet{zhao13}. (9)  \citet{gravity22,gaia23}. Cartesian coordinates with the center at position of the Sun assumed. The $x$-axis points toward the Galactic center, the $y$-axis in the direction of the motion of the Sun,  and the $z$-axis forms a right-handed coordinate system with the other two axes.  }
\end{table}

The paper is organized as follows. In \sect{obs},  we list the various observational constraints on the origin of Fornax and its GCs. We find that Fornax is most probably a tidal dwarf galaxy formed  in an MW-M\,31 encounter 7.5\,Gyr ago; its GCs were probably born in the MW, and we infer the probable orbital initial conditions of the GCs.
In \sect{sim}, we describe our simulations setup and numerical methods. In \sect{res}, we make use of the simulation results to demonstrate that the probability that none of the five GCs of Fornax  sinks in the core is about 20\%.  Moreover, in \sect{other} we find that the simulations explain other observational facts about Fornax: its velocity dispersion (\sect{disp}), the spatial distribution of its GCs (\sect{gcdist}), the origin of its stellar halo (\sect{halo}), and possibly also of the stellar shells in it (\sect{shell}). Moreover, the simulations predict that Fornax has a faint stream (\sect{stream}), and that faint satellites of giant galaxies are lopsided (\sect{lopsided}). Finally, in \sect{stir} we demonstrate that  Fornax might have been transformed from a rotating disk to a nonrotating spheroidal under the gravitational influence of the MW. Our conclusions and main results are summarized in \sect{sum}. In \app{angmom}, we describe the problems we encountered with angular momentum conservation in the simulation code.  Throughout the paper we adopt the MOND acceleration constant $a_0 = 1.2\times 10^{-10}\,$m\,s$^{-2}$. The adopted parameters of Fornax, its GCs, the MW, and the observing frame are listed in \tab{obs}.

\section{Constraints from the observations: A scenario of the formation of Fornax in MOND}
\label{sec:obs}

{During the initial work on this project, it turned out that if we want to answer the question of whether the presence of GCs in Fornax agrees with MOND, we have to take into account the wider context -- in particular, how Fornax and its GCs were created and how they lived. This provides us with the necessary information on how to set up the simulations, how to interpret them, and how to understand their limitations. We infer this information in this section from the analysis of the observational data and partly by making use of older simulations. At the end of the section, we moreover perform additional sanity checks of the proposed formation scenario of Fornax, where we confront its implications with observations and find no conflict.  We progress to the description of our simulations themselves in \sect{sim}.}

\subsection{Fornax is an old tidal dwarf galaxy}
\label{sec:tdg}
The MW is surrounded by the Disk of Satellites (DoS), a planar rotating structure consisting of many, if not all, of its satellite galaxies \citep{pawlowski13vpos}.  The most luminous satellites tend to be located near the midplane of the DoS \citep{banik18}. Fornax is a clear member of this structure, near its midplane, co-rotating with the other clear members of the DoS \citep{pawlowski20}. The standard way to explain the formation of this structure within the MOND framework is that the DoS was formed as a consequence of the past flyby of the MW and Andromeda (M\,31) galaxies. The flyby is an inevitable consequence of the theory \citep{zhao13,kroupacjp,bil21b}. Its detailed simulations show that it is prone to the formation of tidal tails in the galaxies \citep{bil18}. It is possible to tune the orbital parameters within the limits of observational uncertainty such that the tidal structures resemble the DoS and an observed analogous feature in M\,31 \citep{banik18,banik22b}. All these works assume that the members of the DoS are tidal dwarf galaxies (TDGs), that is objects formed by the local  gravitational collapse of tidal tails. Hereafter we assume this origin of Fornax, too. The most elaborate model of the MW-M\,31 encounter \citep{banik18b} found that it happened 7.5\,Gyr ago (i.e., $z\approx 0.9$), which we adopt here. We also assume that Fornax formed from the material of the MW, as indicated by the simulation \citet{bil18}.  We note that simulations of the MW-M\,31 flyby explain also other unusual features of these galaxies \citep{bil18}. Galaxies that are currently similar to the MW have assembled typically one half of their stellar mass at $z=0.9$ \citep{leitner12}.

\subsection{GCs of Fornax originated in the MW}
\label{sec:gcsmw}
The stellar ages of the GCs of Fornax are about 12\,Gyr \citep{deboer16}, that is, they are older than Fornax itself (\sect{tdg}). While it is difficult to determine the exact age of an old stellar population, we assume that it was estimated correctly. The solution to this apparent contradiction turned out to be essential for solving the problem of the survival of the GCs of Fornax in MOND.

The formation of GCs in general is not yet fully understood. One of the main considered options is that at least some GCs are formed from the massive gas clumps and star clusters that are observed in the disks of high-redshift  galaxies \citep{kruijssen15,forbes18,bekki19, kruijssen25}. We assume here that this was the origin of the GCs of Fornax. It is indeed likely that the young MW contained many gas clumps.  For example, the gravitationaly lensed ``Cosmic Snake'' galaxy \citep{cava18}, which has a stellar mass and redshift similar to the MW at the time of the encounter, contains at least several tens of very massive star-forming clumps. They then form massive star clusters (see \citealp{kruijssen25} for a dedicated review). Once the tidal tail of the MW collapsed into the TDGs, the massive star clusters became parts of the satellites as the GCs. An observational image of a similar event was recently published by \citet{whitaker25}. It shows two interacting galaxies at $z=2.53$ with GC candidates concentrated along the tidal tail. We note that if there were any GC in the halo of the MW before the encounter, they could hardly be captured by the forming TDG, since the GCs orbit the MW with a velocity that is much higher than the escape velocity from a TDG.

\begin{figure}
        \resizebox{\hsize}{!}{\includegraphics{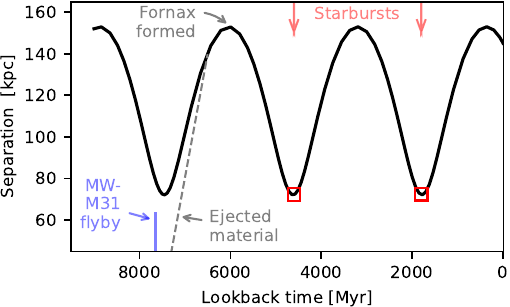}}
        \caption{Separation of Fornax and the MW as a function of the lookback time. We indicate the time of the MW-M\,31 flyby \citep{banik18} that we assume ejected the material forming Fornax. We also indicate our assumed moment of the formation of Fornax, which is the starting point of our simulations, just as the times of the two recent observed peaks of star formation in Fornax \citep{rusakov21}. The orbit of Fornax was not tuned to have apocenters at the observed starbursts peaks.
        } 
        \label{fig:orbit}
\end{figure}

\begin{figure}
        \resizebox{\hsize}{!}{\includegraphics{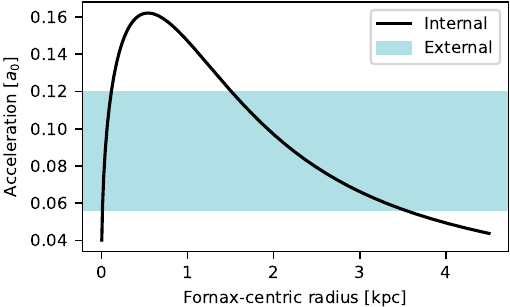}}
        \caption{Internal and external gravitational accelerations in Fornax. The black curve indicates the profile of internal acceleration of Fornax calculated as if it were isolated. The horizontal band indicates the range of the external acceleration experienced by Fornax as it orbits the MW. 
        } 
        \label{fig:accel}
\end{figure}

\subsection{Orbit of Fornax}
\label{sec:orbit}
{Another set of important facts about the life of Fornax comes from the analysis of its trajectory around the MW.  
We calculated it making the use of} the two-body force formula in MOND \citep{milg94c,zhao13}. We used parameters of Fornax and the MW from \tab{obs}. The distance between the MW and Fornax as a function of time is plotted in \fig{orbit}.   Fornax oscillates between about 80 and 160\,kpc, and is leaving its apocenter  presently.  The observational uncertainty of the orbit is explored in \app{unc}.

\subsubsection{Gravitational perturbations by the MW}
\label{sec:pert}
The tidal force competes with dynamical friction by pulling the GCs away from Fornax. Applying the formula by \citet{zhao05}, the tidal radius varies between 2 and 4.5\,kpc. The minimum value is comparable to the projected distance of the most distant GC (\tab{obs}). Because the GCs experience dynamical friction, they were probably further away in the past. The simulations of GC survival thus should include the MW.

In MOND, the external field effect (EFE), stemming from the nonlinearity of its equations \citep{milgmondlaws}, should be considered. In general, the EFE brings the internal gravitational force of galaxies closer to the Newtonian behavior (making them appear more deficient of dark matter  from the perspective of the Newtonian gravity), and causes deformations of the shapes of the galaxies \citep{satelliteefe,qumond,thomas18}. The EFE occurs when the internal gravitational acceleration is smaller than the external gravitational acceleration and the MOND acceleration constant $a_0$. In \fig{accel} we show by the black curve the profile of the internal acceleration of Fornax, calculated as if the galaxy were isolated \citep{qumond}. The range of external accelerations that it experiences during the course of its orbit is indicated by the blue band. The figure shows that the EFE is not negligible in the center and outskirts of Fornax and that it would affect the GCs if they were further away in the past. This is another reason why the MW should be included in the simulations if we want to investigate the survival of the GCs.

\subsubsection{The formation of Fornax}
\label{sec:form}
Interestingly, \fig{orbit} shows Fornax was in the pericenter at about the time when the MW-M\,31 encounter is expected to have taken place. The blue mark in the figure indicates the estimate of the most detailed model of the encounter so far \citep{banik18}: 7.5\,Gyr ago. Within the uncertainties of the current motion of Fornax and the Sun, this third last pericenter of Fornax happened $7.4\pm0.3$ Gyr ago (\app{unc}). 

{We can easily build this in our picture of the life of Fornax.} After the MW-M\,31 encounter ejected the material to form Fornax 7.5\,Gyr ago, the material first moved under the combined gravitational field of the MW and M\,31 (indicated schematically in \fig{orbit} by the gray dashed line). Once M\,31 receded enough, Fornax continued to move along the trajectory we calculated before. It is known from observations and simulations that TDGs start to form very shortly after the ejection, within a few hundreds of megayears at most \citep{lelli15,fensch16,lisenfeld16,baumschlager19}.  The formation of a TDG is initiated by the cooling and gravitational collapse of the gas component \citep{wetzstein07,baumschlager19}. This creates a seed whose gravity gradually attracts more and more material from its surroundings.

As the core of the forming Fornax receded from the MW, its tidal radius increased, such that it could accrete more and more material. The peak of accretion probably happened when Fornax reached its first apocenter about 6\,Gyr ago. We consider this to be the time of Fornax formation. Later, tidal forces started removing material from Fornax -- first the remote parts that have just started moving toward Fornax, later the more inner parts. The question of whether the observed presence of GCs in Fornax contradicts MOND thus means that we have to answer whether they could survive in the galaxy for about 6\,Gyr. 

\subsubsection{Explanation of the peaks in the star formation history of Fornax}
\label{sec:peaks}
{We can add more details life story of Fornax by comparing its orbit and star formation history.} Its detailed reconstruction by \citet{rusakov21} shows two prominent recent peaks 1.8 and 4.6\,Gyr ago. This agrees excellently with the times of the first and second last pericenters of Fornax that we calculated independently (see \app{unc} for the evaluation of their uncertainties). It is indeed credible that the peaks of star formation would occur when Fornax is in pericenter because the tidal forces would deform the circular trajectories of gas clouds, making them collide and form stars. This suggests that we calculated the orbit of Fornax correctly. The match has already been noted and interpreted in the same way by \citet{rusakov21} in the context of Newtonian gravity. For them, however, the match was not so clear because the orbit of Fornax depends on the uncertain parameters of the MW dark matter halo. In MOND, we can easily interpret the initial extended intensive period of star formation in Fornax observed by \citet{rusakov21} as the stars that formed in the MW disk before the encounter with M\,31.

\subsection{Initial orbital conditions of GCs around Fornax}
\label{sec:inigc}
Here we deduce from our scenario of the formation of Fornax what should be the initial orbital conditions of its GCs to be entered in the simulations. It is credible that most of GCs were accreted by the forming Fornax when it was in the first apocenter 6\,Gyr ago. This is when its tidal radius was the largest and the galaxy had the opportunity to accrete material from the largest volume. As an approximation, we will thus assume in the simulations  that at this time, all GCs were in their apocenter with respect to Fornax and started falling on it. 

As for the initial distances of the GCs, there is an upper limit due to the tidal forces. It is given roughly by the tidal radius, even if we detail in \sect{sim} that this estimate is less precise than what one might expect. The point to make here is that it is not possible to solve the problem of the survival of GCs of Fornax simply by postulating that the GCs started their lives far enough from Fornax.

Another limit on the initial spatial distribution of GCs stems from the fact that the GCs initially had to be confined within the mother tidal tail. We reviewed the literature and found that the thickness of the stellar component of the tidal tails that are at least a few tens of kiloparsec long, is around 10\,kpc \citep{wetzstein07,renaud16,baumschlager19,sola22}, both according to simulations and observations. This includes also the tails formed in the simulation of the MW-M\,31 encounter by \citet{bil18}, and the tidal tail with embedded GC candidates observed by \citet{whitaker25}. Therefore, we assume that the GCs started their lives within a cylinder of radius of 5\,kpc centered on Fornax.

As for the initial eccentricity of the orbits of GCs accreted by a TDG, we are not aware of any dedicated study.  We instead use an analogy. Let us first note that a young tidal arm of an interacting spiral galaxy shows a gradient of velocities -- the parts of the tails that are the furthest from the mother galaxy initially recede from it the fastest and those that are the closest recede the slowest. The TDG then forms by the local gravitational collapse of such a structure. The situation resembles the well-studied collapse of dark matter halos in an expanding universe. The radial profiles of the anisotropy parameter of dark matter particles, subhalos, and stars have been studied by \citet{he24}. All these kinematic tracers show qualitatively the same behavior: the orbits are nearly isotropic and the radial anisotropy increases toward larger distances. Based on this, we assume that after the formation of Fornax, the GCs which were located close to its center could be on any shape of orbit, while those further away could only be on almost radial orbits. The only uncertain thing is that we are not able to say beyond which the radius the circular orbit is disfavored. In the simulations, we first do not limit the size of the circular orbits and then we discuss what would happen if their initial radii were limited. 

 \begin{figure}
        \resizebox{\hsize}{!}{\includegraphics{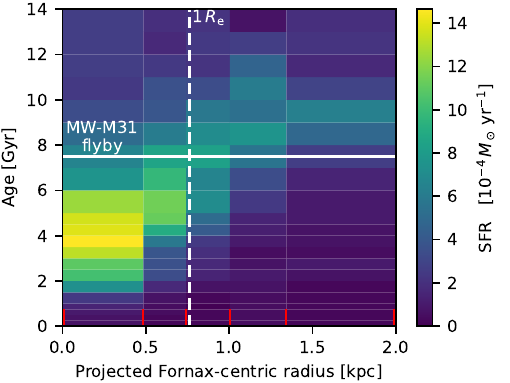}}
        \caption{Spatially resolved star formation history of Fornax. The data come from \citet{deboer12}. The color indicates the star formation rate in the given radial and age bin. Each radial bin contains an almost equal number of stars. The edges of the radial bins are marked by the short red lines. The vertical dotted line marks the effective radius of the galaxy, the horizontal full line the assumed time of the MW-M\,31 flyby \citep{banik18}, when the material of Fornax left the MW in our scenario.
        } 
        \label{fig:sfh}
\end{figure}

\subsection{Fornax was initially gas rich and SN explosions potentially helped the GCs to survive}
\label{sec:gasrich}
Let us complete our picture of the formation of Fornax by assessing its star formation history by \citet{deboer12}. Unlike the recent very detailed star formation history by \citet{rusakov21} discussed above, it covers a substantial part of the galaxy, not only the central regions. It is reproduced in \fig{sfh}.

We note here that a substantial fraction of the stars in Fornax formed during the TDG phase.  The star formation continued until very recently. There are stars that are even only 100\,Myr old \citep{deboer13}. We thus observe Fornax in a very special period of its life, just after quenching. It is even possible that some off-centered gas is still present in the galaxy \citep{bouchard06}. This is very relevant for the problem of the survival of the GCs of Fornax. The simulations of \citet{bil24} showed that the GCs of gas-rich dwarfs in MOND are receiving random pushes because the SN explosions cause large fluctuations of the gravitational potential. This can prevent the GCs of a galaxy from sinking in the center of the galaxy if it is sufficiently gas rich.

From the data in \fig{sfh} we found that since 7.5\,Gyr ago, when the material to form Fornax was ejected, about 70\% of the current-day stellar population of Fornax formed. This means that the initial gas fraction was 70\%, or more if some gas was lost, for example, by ram pressure stripping. For comparison,  at $z=1$ a galaxy that currently has a stellar mass similar to the MW, contains about the same mass in stars, atomic gas \citep{chowdhury22}, and molecular gas \citep{carilli13}, suggesting a total gas fraction of 66\%. The gas fraction in galaxies typically increases with galactocentric radius up to 100\%.  This leaves the role of SNe in the GC survival of Fornax a substantial factor.

In this initial work on the subject, we perform only gasless $N$-body simulations. Instead, the influence of star formation is only discussed on the basis of these arguments.  We note that Fornax was forming stars vigourously to 2-4\,Gyr ago.  We assume that up to this point, the influence of SNe was sufficient to push the GCs away from the Fornax center, but only not farther than 0.5-1\,kpc, where star formation was happening according to \fig{sfh}. We assume that from 2\,Gyr ago the evolution of the GCs would be the same as in our gasless $N$-body simulations.

It is worth mentioning that the star formation histories agree excellently with the TDG formation scenario of Fornax also in other regards.  In \fig{sfh}, we can see that the stars that formed before the MW-M\,31 encounter can be found anywhere in the galaxy and the young ones only in the center. Indeed, gas can dissipate energy, concentrate in the center, and form stars there. In contrast, the dissipation does not act on the pre-existing stars and thus their distribution must remain extended.  Some of the old stars might also have been moved away from the center of the galaxy by the SNe of the central starburst, as in the simulations of \citet{riggs24}.  Interestingly, the chemical models of Fornax by \citet{deboer12} found that gas pre-enriched to a relatively high metallicity had to feed the extended star formation of the galaxy. This is expected if the gas in Fornax originated from the MW.

\subsection{Sanity checks of the scenario}
\label{sec:sanity}
The idea of the formation of the satellites of the MW as TDGs is not fully established in the community yet. To our best knowledge, the suggestion that their GCs are pre-existing MW star clusters appears here for the first time. These proposals might raise various doubts. Before proceeding toward the description of our simulations, we make here several sanity checks of them.

\subsubsection{Could Fornax be a TDG?}
\label{sec:sanitytdg}
\citet{duc14} argued against the DoS satellites being TDGs, because the observed TDG candidates have larger radii at a given stellar mass than the satellites. This discrepancy is, however, not present in newer and bigger samples of TDGs \citep{ren20,zaragozacardiel24}. 

In our scenario, Fornax was initially gas-rich TDG. All observed young gas-rich TDGs show the kinematics of rotating disks \citep{lelli15}. The current Fornax is, however, not a disk and does not show any appreciable rotation \citep{walker06,martinezgarcia21}. We address this issue in \sect{stir}, where we make a dedicated simulation which demonstrates that the gravitational influence of the MW itself is capable of causing the necessary transformation.

Fornax has a very relaxed shape of its outer isophotes \citep{bate15,yang22}. This might appear surprising for a TDG which formed in a cataclysmic even not so long ago. Nevertheless, there are very few stars observed beyond 2.5\,kpc \citep{yang22}, such that any irregularity of the galaxy would be difficult to recognize. Below  2.5\,kpc, any tidal structures would be erased by the phase mixing mechanism \citep{mo}:  at 2.5\,kpc, even if we assume the external field in perigalacticon, material on a circular orbit would complete over seven revolutions since the formation of Fornax 6\,Gyr ago. The images in \citet{duc14} reveal that the 4\,Gyr old TDGs in NGC\,5557 already have a rather relaxed appearance. Already these images of not-so-old TDGs also show that the mother tidal tails, from which the TDGs formed, are much fainter than the dwarfs themselves \citep{duc14}, so they would be difficult to notice for Fornax. The detailed mechanism is illustrated by the images of the simulations of \citet{wetzstein07}. They show that as time progresses, the mother tidal tail appears as a longer and longer band wrapped around the mother galaxy. As the material becomes more and more diluted, it becomes unobservable.

\subsubsection{Could the GCs of Fornax have formed in the MW?}
\label{sec:sanitygc}
In our scenario, the GCs of Fornax were initially massive star clusters in the MW. If the TDG origin of Fornax is true, then similar GCs should be present also in other massive members of the DoS. Indeed, Fig.~9 of \citet{pace21} reveals that the Magellanic Clouds and the Sagittarius dwarf all contain GCs with ages and metallicities similar to those of Fornax.

Also, images of confirmed old TDGs indicate that not all material in a tidal tail collapses into TDGs \citep{duc14}. Therefore some of the GCs formed in the MW should also be in the region of the DoS in between of the satellites. Again, this is the case: \citet{pawlowski13vpos} found that the so-called young halo GCs and some stellar streams concentrate in the DoS, forming together the ``Vast Polar Structure''. \citet{mackey04} found that the properties of the GCs of the satellites within the DoS agree with the properties of the young halo GCs, suggesting their common origin. On the contrary, these GCs differ from the old halo and bulge GCs populations of the MW, which probably formed differently. 

Next, one might wonder why none of the ancient massive star clusters survived within the MW disk, but they did in the supposed TDGs as the GCs. The survival of massive star clusters was discussed in detail in the review \citet{kruijssen25}. It turns out that if even a very massive star cluster does not migrate away from its birth place, it is destroyed by repeated tidal interactions with giant molecular clouds. If the GC gets into a TDG, it does not have problems to survive, as evidenced by the observations of old GCs in gas-rich dwarfs \citep{sharina05,georgiev09,bil24}.

Finally, we might wonder why no new GCs formed within Fornax, given that it was gas-rich and TDGs form stars intensively. However, dedicated studies found no convincing evidence for the formation of long-lived GCs in TDGs \citep{fensch19}. From this perspective, it seems reasonable that only the newly discovered young, light-weight, and metal-rich GC Fornax6 \citep{pace21} was formed in the TDG phase.

\begin{table}
\caption{Computational parameters of PoR used in the main set of simulations (Sects.~\ref{sec:sim} and~\ref{sec:res}).}
\label{tab:setup}
\centering
\begin{tabular}{ll}
\hline\hline                   
Parameter  & Value               \\
\hline 
\texttt{levelmin} & 5\\
\texttt{levelmax} & 21 \\
\texttt{boxlen} & 409.6\,kpc \\
\texttt{mass\_sph}  & 1\,$M_\sun$ \\
\texttt{m\_refine} & 16000\,$M_\sun$ \\
Maximum resolution & 0.2 pc \\

\hline                                            
\end{tabular}
\end{table}

\begin{table*}
\caption{Statistics of sunk, surviving, and escaped GCs for different assumptions. }
\label{tab:results}
\centering
\begin{tabular}{l|lll|lll|lll|lll|l}\hline\hline
& \multicolumn{6}{c|}{Light GC}&  \multicolumn{6}{c|}{Massive GC} & \\\cline{2-13}
Layer &  \multicolumn{3}{c|}{Radial}&  \multicolumn{3}{c|}{Circular}&  \multicolumn{3}{c|}{Radial}&  \multicolumn{3}{c|}{Circular} & $\gamma$\\\cline{2-13}
[kpc] & $n_c$      & $n_s$     & $n_e$ & $n_c$      & $n_s$     & $n_e$ & $n_c$      & $n_s$     & $n_e$ & $n_c$      & $n_s$     & $n_e$     & [kpc$^3$] \\\hline
0-1   & 10$^*$ & 0$^*$ & 0$^*$ & 9.95 & 0.05 & 0    & 10$^*$ & 0$^*$ & 0$^*$ & 10$^*$ & 0$^*$ & 0$^*$ & 2.1      \\
1-2   & 10     & 0     & 0     & 2.54 & 6.46 & 1    & 10$^*$ & 0$^*$ & 0$^*$ & 10     & 0     & 0     & 15       \\
2-3   & 1.77   & 7.23  & 1  & 0    & 2.98 & 7.02 & 10     & 0     & 0     & 4.61   & 3.39  & 2     & 40       \\
3-4   & 0.22   & 3.42  & 6.37  & 0    & 0.78 & 9.2  & 8.73   & 0.27  & 1     & 0      & 1.1   & 8.9   & 77       \\
4-5   & 0.02   & 4.98  & 5     & 0    & 0.17 & 9.83 & 2.76   & 2.24  & 5     & 0      & 0.39  & 9.61  & 130      \\
5-6   & 0      & 0.83  & 9.17  & 0    & 0.55 & 9.45 & 0.05   & 2.9   & 7.05  & 0      & 0.29  & 9.71  & 110      \\
6-7   & 0      & 0     & 10    & 0    & 0    & 10   & 0      & 0.76  & 9.24  & 0      & 0     & 10    & 96 \\\hline 
Survival probability &  & 0.83 &  &  & 0.86 &  &  & 0.30 & & & 0.46 &  \\\cline{1-13}
\end{tabular}
\tablefoot{The ``Layer'' column gives the radial range within which the GCs start. Ten GCs start from each layer. The ``Light (Massive) GC'' refers to the $1 (5)\times 10^5\,M_\sun$ GC. The columns  $n_c$,  $n_s$, and $n_e$ show the numbers of sunk (as for ``core''), surviving, and escaped GCs in each layer, respectively. The asterisk symbol indicates that the numbers are assumed, without actually making the simulation.   The column $\gamma$ lists the geometric weights of the layers. }
\end{table*}

\section{Simulations}
\label{sec:sim}
The main objective of this paper is to investigate whether the observations of the five GCs in Fornax excludes QUMOND, using high-resolution $N$-body gas-less simulations with a spherical Fornax. Unless specified otherwise, each simulation contained the MW, Fornax and one GC.

We did the simulations using the public adaptive-mesh-refinement code Phantom of Ramses (PoR, \citealp{Lughausen_2015,Nagesh_2021}), which is a patch to the RAMSES code of \citet{ramses} that can solve the QUMOND equation for gravitational field \citep{qumond}. We used the 
so-called ``simple'' interpolation function $\nu(x) = \frac{1}{2} + \sqrt{\frac{1}{4} + x^{-1}}$. The setup of the numerical solver of PoR is summarized in \tab{setup}.

The MW was modeled by a S\'ersic sphere of the index 0.5, mass of $6\times 10^{10}\,M_\sun$ (\tab{obs}) and effective radius of 3\,kpc. However, the exact inner structure of the MW model does not play a substantial role because the minimum distance of Fornax from the MW is much greater than the effective radius of the MW. The {\sc phantom\_staticparts} patch of PoR \citep{Nagesh_2021} was used to achieve that the MW particles do not move with respect to the computational grid so to save computational time. The galaxy consisted of $10^3$ particles.

Fornax was modeled as a S\'ersic sphere with the observed parameters (\tab{obs}). It consisted of $1.5\times10^6$ moving particles, each of 20\,$M_\sun$. The model was prepared using the method described in \citet{Bilek_2022b}. In short, for a prescribed density profile, and anisotropy parameter, the spherical Jeans equation is solved to get the radial profile of velocity dispersion. Velocities of particles are then drawn from a Gaussian distribution according to this solution and the assumed anisotropy parameter, which was set here to be zero. The model is then let evolve for 8\,Gyr in isolation in order to establish an equilibrium. 

To find the initial orbital parameters of Fornax with respect to the MW, first we analytically integrated their orbit backward until they reached an apocenter. We oriented the simulation Cartesian coordinate system such that: 1) the MW is in the origin, 2) Fornax initially lies on the $x$-axis, 3) it moves in the direction of the $y$ axis, and 4) the $z$ axis complements a right-handed coordinate system. These settings represent the observed Fornax at the epoch of its supposed formation 6\,Gyr ago (\sect{form}).

\begin{figure*}[]
        \centering 
        \includegraphics[width=17cm]{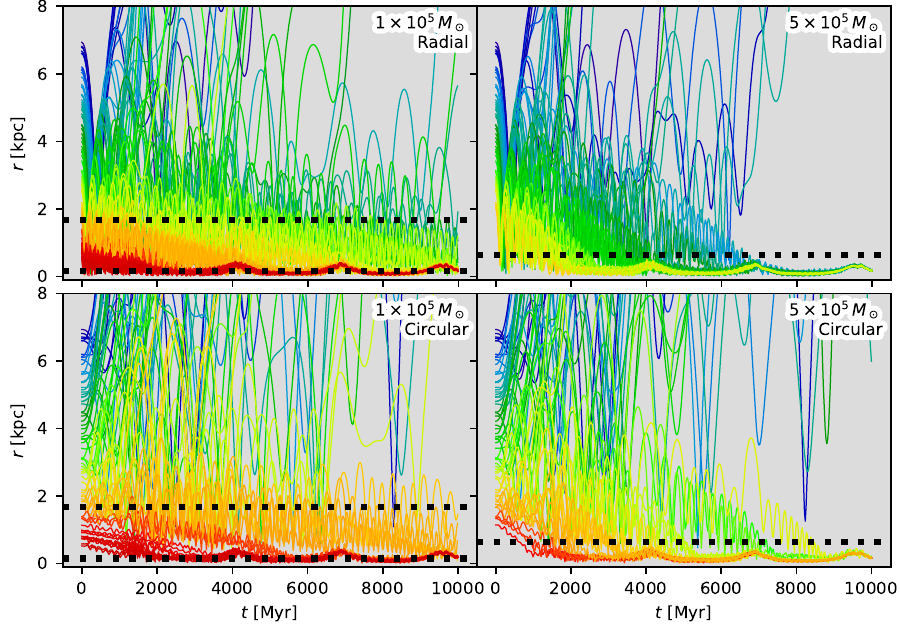}\vspace{-1ex}
     \caption{Orbital decay of the GCs of Fornax, depending on their mass and initial trajectory. The vertical axis denotes the distance between the GG and the center of Fornax. For the light GCs, the two dotted horizontal lines indicate the range of the projected distances of the four lightest GCs of the real Fornax.  For the massive GC, the horizontal dotted line indicates the projected distance of the most massive GC of the real Fornax.}
        \label{fig:tracks}
\end{figure*}

We initially intended to model the GCs as resolved multibody objects. However, the simulations did not conserve the angular momentum, which made the GCs sink unrealistically quickly. As a warning, we give more details on these simulations in \app{angmom}. We then continued by simulations where the GCs were represented by point masses. These simulations conserve the angular momentum much better. The potential disadvantages of representing the GCs by point masses is that the nonzero size of the object influences the dynamical friction somewhat \citep{white76,mo}  and that the GCs can experience extra dynamical friction by interacting with each other and even merge \citep{duttachowdhury20, bil21}. Tides on the GCs from Fornax are negligible: the tidal radius is at least a few tens of parsecs anywhere in the galaxy.

The GCs had either $1\times 10^5\,M_\sun$ or $5\times 10^5\,M_\sun$. The former is the typical observed mass of the four lightest GCs in Fornax and the latter the maximum mass (\tab{obs}). We refer to them below as the light and massive GCs. The GCs were initiated in the apocenters of either their orbits for the reasons explained in \sect{inigc}. We explored only the exactly circular and exactly radial orbits. For the circular orbits, the initial velocity of the GC with respect to Fornax was estimated using the \citet{freundlich22} formula for the MOND acceleration in the presence of an external field. In each simulation there was only one GC. In \app{multigc}, we show that including all five GCs in the simulations does not affect our conclusion substantially.  

In our initial experiments, we arranged the GC to start at one of the coordinate axes at different distances and watched if it can remain in the galaxy unsunk for 6\,Gyr\footnote{As a side-note, we briefly mention a noteworthy fact that we noticed in these initial experiments. Contrarily to the ``common knowledge'', when we put the GC beyond the tidal radius of Fornax with a zero velocity with respect to it, the GC was not stripped. Instead, it started falling on Fornax, and was stripped only later. Actually, the fact that the tidal radius is just an approximate concept has been realized a long time ago already \citep{henon70,bt08}.}. It turned out that the results are very sensitive to initial conditions, such that it is hard to judge how probable the observation of the five GCs in Fornax is. This agrees with the well-known chaotic nature of the three-body problem, whose form we are dealing with here.

We hence took a statistical approach. The space around Fornax was divided into spherical layers, each 1\,kpc thick (see the details in \tab{results}). For each layer, we run ten simulations, in each of which the GC started at a random position and, in the case of the circular orbits, with a randomly oriented velocity vector. Then we could estimate how probable the observed state of GCs in Fornax is in MOND.

\section{The survival of the GCs}
\label{sec:res}
In order to explain the observed presence of the GCs in Fornax, our simulations should show the they can survive there for 6\,Gyr, that is the time since the assumed formation time of Fornax \sect{form}. This, however, holds true only if we ignore the facts that Fornax was forming stars in its center until  2\,Gyr ago (\sect{gasrich}), and they could have hold the GCs away from the center of Fornax by the SN explosions \citep{bil24}. Below, we first progress like the hydrodynamical effects played no role. Then we discuss how the results would probably change if the effect of SNe were included.

In \fig{tracks}, we plotted, for all simulations together, the distances of the GCs from the center of Fornax as function of time. The center of Fornax was determined by the sigma-clipping algorithm, in order to exclude the particles that escaped from Fornax to form a tidal stream (\sect{stream}). In the tiles with the massive GCs, the  projected distance of the most massive GC of the real Fornax is indicated by the dotted horizontal line. For the light GCs, the dotted horizontal lines indicate the minimum and maximum observed projected distances of the four lightest GCs of the real Fornax.

The GCs show three types of orbits in the simulations. First, they are the escaping orbits. The particles that started far away from Fornax escape the easiest, but also the concrete location and velocity influence whether the particle escapes. Many escaped particles receded beyond 100\,kpc from Fornax within a few Gyrs. 

Then we have the sinking orbits, at which the GC sinks in the center of the galaxy due to dynamical friction. These are followed primarily by the GCs released from the inner layers. The small peaks at these orbits at around 4, 7 and 10\,Gyr are caused by the deformation of Fornax by tides and the EFE from the MW (\sect{lopsided}), such that Fornax does not have a well-defined center.  Some deformation is present even in the apogalacticon. Nevertheless, we can see that if the GCs started their lives at the currently observed positions, they would most probably sink in a few Gyr. This confirms the older analytic results \citep{ciotti04, sanchezsalcedo06, angus08}. The core stalling effect starts being efficient below ca. 0.3\,kpc, comparable to the projected distance of the nearest observed GC. This agrees with the simulations of GCs of ultra-diffuse galaxies \citep{bil21}, where it was encountered only below half of the effective radius too. 

Finally, we have the intermediate surviving orbits, which do not escape nor sink.  The classification obviously depends on the time in question. On very long timescales, all GCs either sink or escape.  Importantly, \fig{tracks} already shows the answer to the main objective of this paper: in MOND, it is not impossible to observe surviving GCs in Fornax, even without SNe.

Nevertheless, this does not tell us how probable the observed state is, namely that Fornax has five unsunk GCs, out of which one has $5\times10^5\,M_\sun$ and the other four about $1\times10^5\,M_\sun$.  In the rest of the section, we estimate this probability.

We first needed to opt for an exact definition of what should be called a sunk and an escaped GC. We chose an approach that mimics the observations. We inspected the projected distances of the GCs as they would be observed by an observed in the center of the MW. Using this position of the observer rather than at the position of the Sun makes only a negligible difference because Fornax is at least eight times further from the MW than the Sun. We note that the projected distance, unlike the three-dimensional, is immune to the deformations of Fornax, because they happen in the direction toward the MW.  At a given time, we conservatively considered a GC sunk if its projected distance from the center of Fornax was smaller projected distance of the innermost observed GC of the real Fornax, that means 0.15\,kpc (\tab{obs}). We note that with this definition, for example, a GC that oscillates along the line-of-sight is always considered sunk. At a given time, we considered a simulated GC to be escaped if its projected distance from Fornax was greater than 3\,kpc. The limit was set in order not to consider a GC surviving if its distance is much greater that the distances of the observed GCs. In particular we set the limit to be three times the average projected distance of the observed GCs from Fornax. We note that with this definition, a GC can be considered surviving even if its distance from Fornax along the line-of-sight is very large. The particles that were not sunk or escaped at a given time were considered surviving and counterparts of the GCs of the real Fornax.

Ideally, we would count the number of sunk, escaped, and surviving GCs at the time of 6\,Gyr after the simulation start. The simulated GCs however oscillate around Fornax at a relatively short timescales (\fig{tracks}). Shortly before or after the time of our interest, the classification of a GC might change. Figure~\ref{fig:tracks} shows that this might change the results substantially, mostly for the massive GCs and circular orbits, since there are just a few candidates for surviving  GCs at 6\,Gyr. For this reason, each GC was counted in each category as a fraction, according to the fraction of time spend in the category within the period  5.5-6.5\,Myr. For each layer within which GCs are launched,  we then added up the counts of GCs in each category. The results are stated in \tab{results}. Here,  $n_s$ stands for the surviving category, $n_c$ for sunk (they are in the {\bf C}ore),  and $n_e$ for the escaped. These counts can be thought as proportional to the average number of GCs  that will have the given destiny if launched from a unit of volume of the given layer. We assumed that all GCs that would start outside of the outermost layer would escape.

{Nevertheless, before calculating the probabilities of the different destinies of GCs, we must take into account the geometry of the initial conditions of the GCs. As discussed in  \sect{inigc}, in our scenario Fornax and its GCs started their life by collapsing from a tidal tail, which we approximate by a cylinder of a radius of 5\,kpc, filled by matter homogeneously. This leas us to assign to each of the layers from which the GCs are launched, a geometric weight $\gamma$, which is defined as the volume of the intersection of such a layer with a cylinder whose axis goes through the center of the layer and has a radius of 5\,kpc. The values of the geometric weights for the different layers are stated in \tab{results}.  We denote by $X$ the number of GCs per each unit of volume of the tidal tail.}

Now we can evaluate the probability that a GC that has not escaped from Fornax will survive. {If $X$ GCs were launched from each unit of  volume $\gamma_i$ of a given layer $i$, we expect the total of $\frac{X}{10}n_{s,i}\gamma_i$ of them to survive.  If $X$ GCs were launched from each unit of the volume of all layers, $N_s = \sum_i  \frac{X}{10}n_{s,i}\gamma_i$ of them would survive and $N_c = \sum_i \frac{X}{10}n_{c,i}\gamma_i$ would sink. From this, we get that the probability that a GC that does not escape will survive is $P_s = N_s/(N_s+N_c)$, which is a number that does not depend on $X$. The concrete probabilities of surviving} for the massive and light GC and radial and circular orbits are stated in the last line of \tab{results}. We can see that the probabilities depend more on the mass of the GC than on the shape of the orbit. The probability of surviving is about 85\% for the light cluster and 40\% for the massive one. We can see from \tab{results} that if the circular orbits with initial radii over 3\,kpc were banned (\sect{inigc}), the probability would almost not change, because the vast majority of such orbits is escaping.

This gives the final probability of the whole observed census of the GCs of Fornax. For the four lighter GCs and one massive GC of Fornax we get the probability of $0.85^4\times0.4 = 20\%$. To fully appreciate this result, we also have to consider that for our study we chose the most problematic satellite of the MW.  The GCs of Fornax thus do not present a problem for MOND.

It is worth noting that in the simulations of \citet{leung20}, the mass of Fornax3 was $3.63\times 10^5\,M_\sun$, that is less than we assumed here. If the mass of the GC is indeed this, then the explanation of its survival becomes even easier.

Now let us consider the effect of the SNe. As explained in \sect{gasrich} we assume that the SN explosions were able to keep the GCs 0.5-1\,kpc from the center of Fornax, but only until at most 2\,Gyr ago. Figure~\ref{fig:tracks} then indicates that the SNe might have helped only the innermost observed GC  to survive, if it was on a circular orbit 2\,Gyr ago. The massive cluster, even if were supported by SNe until 2\,Gyr ago, would had sunk until today anyway. The support by the SNe thus does not appear as the important for the survival of the GCs.

\section{Other interesting results}
\label{sec:other}
\begin{figure}
        \resizebox{\hsize}{!}{\includegraphics{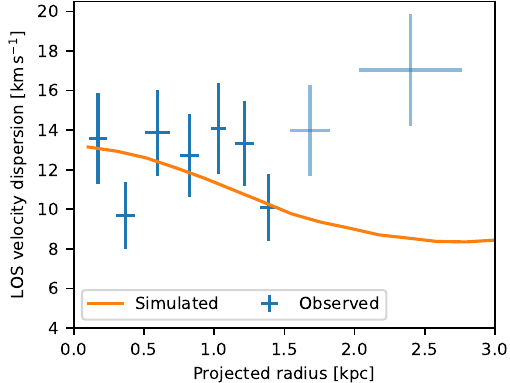}}
        \caption{Comparison of the simulated and observed line-of-sight velocity dispersion profiles of Fornax. The unreliable observational data points are indicated by the fainter color.
        } 
        \label{fig:siglos}
\end{figure}

\subsection{Velocity dispersion profile reproduced}
\label{sec:disp}
In our simulations, we used the stellar mass of Fornax inferred from stellar population models. However, the stellar mass, more precisely its profile, determines the shape of the velocity dispersion profile of the galaxy \citep[e.g.,][]{stigari06,angus09,jardel12,pascale18}. We can then ask whether the simulations reproduced the observed velocity dispersion profile of Fornax. 

To this end, we measured in the simulation the line-of-sight velocity dispersion, as if it were observed from the center of the MW. We measured it in circular radii at the time corresponding to the present day, i.e., at the simulation time of 6\,Gyr.

The observed velocity dispersion profile depends somewhat on what criterion is used to distinguish the stars of Fornax from the foreground MW stars, and how the observed elongation of the galaxy is taken into account. We compared our simulations to one of the measurements by \citet{walker06}, namely the one based on elliptical apertures and 186 stars. The same dataset was analyzed by \citet{serra10}. Their method indicated that all stars further than 1.5\,kpc from the Fornax center are contaminants and thus the velocity dispersion measurements in this region should not be used. 

The comparison of the data to the simulation is shown in \fig{siglos}. The excluded data points are indicated by the lighter color.  There is a good agreement between the simulated profile and the reliable measurements. Our simulations predict that at large radii, the velocity dispersion is lower than currently measured.

The influence of tides on the velocity dispersion profile is rather weak over the plotted range. The profile extracted for the beginning of the simulation is just about 0.5\,km\,s$^{-1}$ higher than the plotted one.

\begin{figure}
        \resizebox{\hsize}{!}{\includegraphics{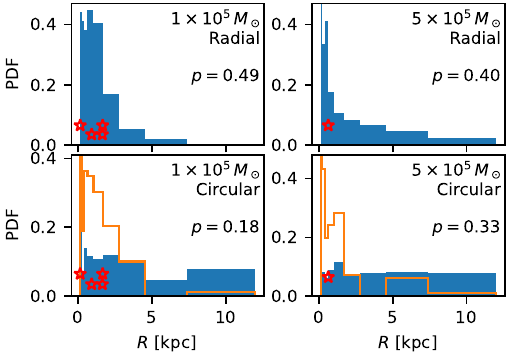}}
        \caption{Comparison of the simulated and observed distributions of the projected distances of the GCs of Fornax. The blue histograms are constructed from all the simulated GCs. In orange histograms, we excluded the circular trajectories with initial radii over 3\,kpc, to take into account the theoretical considerations from \sect{inigc}. The red stars mark the projected distances of the observed GCs. Indicated are the $p$-values of the KS test of the null hypothesis  that the simulated and observed distributions are the same.  For the circular orbits, the $p$-values are calculated for the restricted initial radius. A value below 0.05 means inconsistency.
        } 
        \label{fig:gchist}
\end{figure}

\subsection{Explanation of the observed distribution of the GCs of Fornax}
\label{sec:gcdist}
In this section, we analyze the distribution of the projected galactocentric radii of GCs predicted by our simulation and show that it agrees with the observed distribution.  The expected GC distributions was generated following the logic  from \sect{res}. We used the projected distances as they would be seen by an observed in the center of the simulated MW. We used the trajectories of all simulated GCs from \sect{res}. Each trajectory was weighted by the geometric weight $\gamma$ corresponding to the layer where the GC started. We considered all positions of the GCs between the simulation times 5.5-6.5\,Gyr. We are interested only in the GCs that were not sunk, and thus we had to exclude the GCs whose projected distance is less than 0.15\,kpc. From all measurements of the projected distances, we then created a weighted histogram, which was then normalized in the range 0.15-12\,kpc to unity in order to get a probability distribution function (PDF).

We first applied this to all the simulated GCs. The resulting PDFs are shown in \fig{gchist} by the blue color, for both types of orbits and both GC masses.   For the radial orbits, the PDFs peak at small radii and vanish toward large radii. In contrast, for the circular orbits, the PDFs are rather flat. Figure~\ref{fig:tracks} reveals that the flatness is caused by the GCs which start at large distances from Fornax. As explained in \sect{inigc}, such orbits are disfavored by our formation scenario of Fornax. If we exclude the circular orbits which started more than 3\,kpc away of Fornax, we get the PDFs shown in orange in \fig{gchist}. Their shapes are similar to those of the radial orbits. 

The positions of the observed GCs are indicated by the red stars in \fig{gchist}. We can see that the GCs could not start their lives on circular orbits far from Fornax, in agreement with the theoretical expectations of our scenario.
We then made a two-sided two-sample KS-test to check the consistency of the observed and simulated distributions of projected distance of GCs. We included only the simulated GCs that are closer than 12\,kpc. We also excluded the simulated GCs that are closer than 0.25\,kpc, because the some of PDFs had abrupt strong spikes in this region from the nearly sunk GCs. The $p$-values are stated in \fig{gchist}. They show the consistency of the simulated and observed distributions of GCs. If the additional cut is not imposed, the $p$-values actually indicate even a better consistency. 

\begin{figure}
        \resizebox{\hsize}{!}{\includegraphics{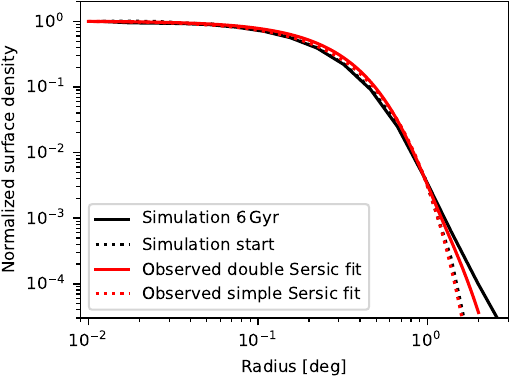}}
        \caption{Simulated Fornax develops a stellar halo similar to the real Fornax. The so-called ``halo'' of Fornax \citep{yang22} is the excess of stars in the outer part of the galaxy with respect to a simple S\'esics fit (red dotted line). The whole profile is described well by a sum of two S\'esic profiles (full red line). The simulated Fornax is initially described by a simple S\'esic profile (black dotted line), but eventually develops a profile with a ``halo'' (full black line) like the real galaxy has.
        } 
        \label{fig:sdprof}
\end{figure}

\subsection{Diffuse stellar halo of Fornax reproduced}
\label{sec:halo}
\citet{yang22} extracted the surface density profile of Fornax up to extremely large radii.  They found that if it is fitted by a S\'ersic profile, the observed profile deviates from it substantially at large radii. They called this outer deviating region a halo of the galaxy. They found that the whole profile of the galaxy can be fitted by a sum of two S\'ersic profiles. 

We were interested whether we can see a similar feature in our simulations. Our simulated Fornax was initialized in order to follow the single-S\'ersic fit by \citet{yang22} (\tab{obs}). Because the simulated galaxy experienced the loss of a few percent of its stars because of  tides and the EFE, the profile had to evolve. We extracted the surface density profile from the simulation at the time of 6\,Gyr as it would be observed from the center of the MW.

The result is compared to the observed profiles in \fig{sdprof}.  Both single and double-Se\'rsic fit of the observed data are shown. For the simulation, we show the surface density profile at the beginning of the simulation at the current time. All profiles were normalized to be one in the center of the galaxy. The figure shows that the observed double-S\'ersic profile deviates upward from the observed single-S\'ersic fit beyond 1\degr. It also shows that the  observed single-S\'esic profile nearly coincides with the initial simulated profile, as intended. Most importantly, after 6\,Gyr of evolution, the simulated profile shows a similar upturn as in observations. The upturn from  the single-S\'ersic fit happens at the same radius as observed and has a similar shape and a correct order of magnitude. 

We note that a better agreement cannot be expected because we used simplified initial conditions. The real Fornax at the beginning of the simulated period, was most probably not an $N$-body sphere but a rotating disk with a substantial fraction of gas, see Sects.~\ref{sec:obs} and~\ref{sec:stir}.

\begin{figure}
        \resizebox{\hsize}{!}{\includegraphics{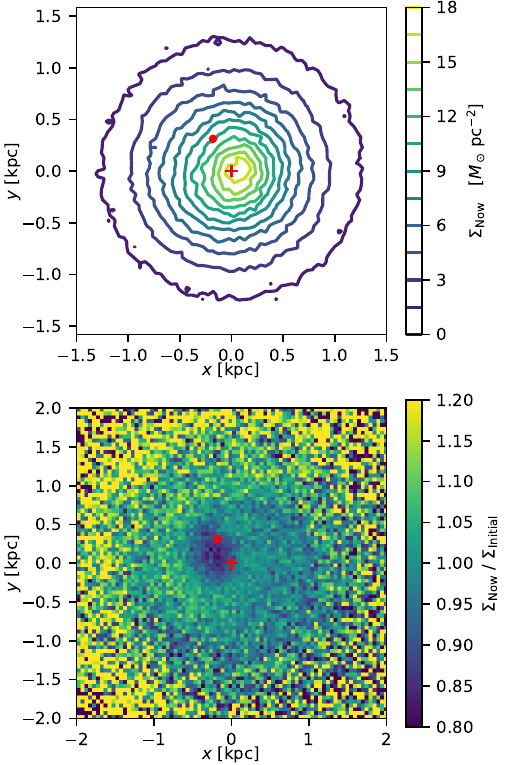}}
        \caption{Induction of substructure in Fornax by a massive GC.  Top: Contours of surface density. The cross indicates the barycenter of the galaxy. The dot indicates the position of the GC. Bottom: Ratio of the current surface density to that at the beginning of the simulation. The figure represents a view along a line perpendicular to the orbital plane of the GC. {Views of the perturbation from other lines of sight are provided in \app{add}.}}
     \label{fig:shell}
\end{figure}

\subsection{The shells in Fornax as a feature provoked by the most massive GC}
\label{sec:shell}

In the simulations involving the massive GCs, we noted that it creates substructures in the surface density maps of the galaxy. It is intriguing that they somewhat resemble the observed shells in Fornax \citep[e.g.,][]{wang19}. These substructures formed in the simulations because of the varying gravitational force exerted by the moving GC on the core of the galaxy. The situation resembles two merging galaxies. The effect is most visible shortly before the sinking completes.

For the purpose of visualization, we run a separate simulation without the MW. This was necessary, because the effect of the massive GC interferes with the deformations of the galaxy by the tides and the EFE from the MW (\sect{lopsided}). The $5\times10^{5}\,M_\sun$ GC was initially placed on a circular trajectory with a radius of 2\,kpc. The perturbations induced in the galaxy by the massive GC are  illustrated in \fig{shell} (see also \fig{shellall}). It shows the simulation at 3\,Gyr. The upper panel shows how the isophotes of the galaxy are deformed. The GC made the core of the galaxy to be offset with respect to the outer isophotes and the barycenter of the galaxy. The core of the observed Fornax is offset in a similar fashion \citep{wang19}. The bottom panel of \fig{shell} shows how the stellar surface density changed with respect to the surface density at the beginning of the simulation, when Fornax was spherically symmetric. This figure emphasizes the substructures that the GC induced in the galaxy. Their size is about 1\,kpc, which corresponds to the angular size of 0.4\degr. This is roughly the size of the observed shells of Fornax \citep{wang19}.

\begin{figure}
        \resizebox{\hsize}{!}{\includegraphics{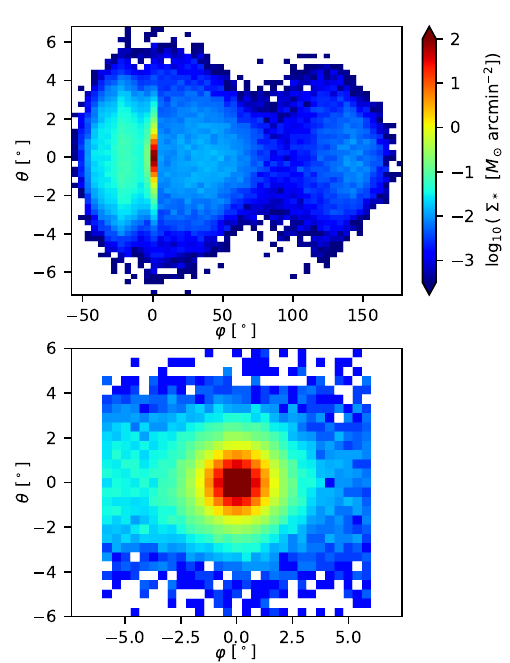}}
        \caption{Simulated view of the predicted tidal stream of Fornax. The upper panel shows the whole stream. The color indicates the surface density of the stars. In these coordinates, Fornax moves to the right. Note that the vertical and the horizontal coordinates do not have the same aspect. The bottom panel only shows the vicinity of Fornax.
        } 
        \label{fig:stream}
\end{figure}

\subsection{Prediction of the stellar stream extending of Fornax} 
\label{sec:stream}
The simulated Fornax developed a stream of stripped particles. No stream extending from the real Fornax has been discovered yet. We address in this section how the predicted stream is supposed to look like, how it is possible that it has not been detected yet, and whether it is observable with the existing instruments.

We constructed a surface density map of the simulated stream as it would appear to an observer in the center of the MW at the simulation time of 6\,Gyr. We constructed it in the angular coordinates $\phi$-$\theta$. The angle $\phi$ was measured in the orbital plane of Fornax. It had its origin in the center of Fornax and increased in its direction of the motion. The angular coordinate $\theta$ was perpendicular. The map is shown in \fig{stream} -- both for the whole stream and for the vicinity of Fornax. 

The deepest observational map of surface density of Fornax known to us is that by \citet{yang22}. They could trace Fornax to about 2.5\degr before the contamination by the MW stars became too high. Their maps do not show the stream. Figure~\ref{fig:stream} shows that it agrees with our simulation, where there is no substantial deviation from a regular shape at this distance. Detecting the deformation of the real Fornax is further complicated by the fact that it is elliptical.

Nevertheless, future observational studies have a chance to detect the stream if they focus on the bright spot in the trailing arm, which is located about 20\degr from the galaxy. Its surface density is predicted to be about the same as at 2.5\degr, that is at the limit of detectability of  \citet{yang22}. As in \sect{halo}, we should point out that these are just rough estimates. The actual surface density can be higher or lower, depending on what was the real mass distribution of Fornax at the beginning of its life, and also because the simulation ignores hydrodynamics.

Finally, we note that in \fig{stream}, the leading part of the stream is longer and fainter than the trailing part.  While the projection effects influence the appearance of the stream in this figure, we note that in three dimensions, the leading part is still longer than the trailing one.  This result is the opposite to the result of \citet{thomas18}, probably because of different orbits of the investigated objects. During the course of our simulation, the leading arm is sometimes longer and sometimes shorter than the trailing arm. This is because of the eccentric trajectory of Fornax, because the stream particles in apocenter move slower than the particles in pericenter. The trailing arm, however, always contains more particles, as in \citet{thomas18}.

\begin{figure}
        \resizebox{\hsize}{!}{\includegraphics{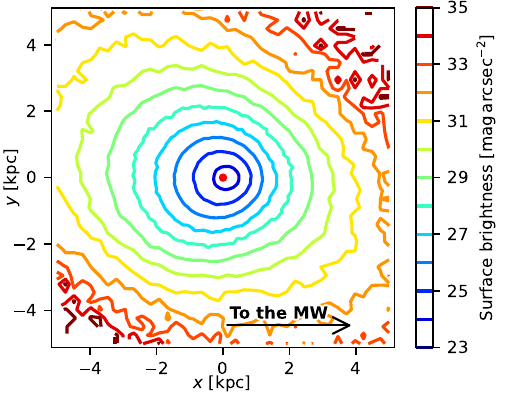}}
        \caption{ MOND predicts the lopsided and ovoidal isophotes of satellite galaxies. The figure shows the surface brightness of Fornax in our simulation projected to the orbital plane in the moment of pericenter.  The absolute magnitude of the Sun in the $V$ band was assumed to be 4.83\,mag \citep{binneymerrifield98}. The red dot indicates the barycenter. The  MW is to the right. 
        } 
        \label{fig:shape}
\end{figure}

\subsection{MOND-specific deformation of shapes of satellite galaxies}
\label{sec:lopsided}
The shape of Fornax in our simulation was deformed because of the gravitational interaction with the MW, see \fig{shape}. Fornax is shown here at the moment of its second pericenter (\fig{orbit}). The isophotes are lopsided in the direction approximately away from the MW, except for the innermost isophotes, which are lopsided in the opposite direction. Moreover, the isophotes have an ovoidal shape with the sharper end pointing toward the MW. Similar effects have already been reported to distinguish MOND from  Newtonian gravity \citep{candlish18,thomas18}. These works have explored them for galaxies falling in galaxy clusters and  GCs orbiting the MW. Here we briefly discuss whether this characteristic signature of MOND could be detected using nearby dwarf spheroidal or elliptical satellites.

The satellites of the MW are not very suitable, because the deformation happened along the line connecting the satellite with the MW. We, as observers located relatively close to the MW center with respect to the distances of the satellites, cannot detect the deformations. It seems  more promising to search for the ovoidal and lopsided isophotes for the satellites of the nearby giant galaxies, primarily the Andromeda galaxy and Centaurus~A. For the Andromeda system at least, it is possible to resolve individual stars within the galaxies, such that we can reach an extremely deep surface-brightness limit. For example, the PAndAS survey detected structures in the stellar halo of Andromeda which are as faint as 32-33\,mag\,arcsec$^{-2}$ \citep{ibata14b}. 
The observations should focus on the satellites for which the line connecting them with their host is close to perpendicular to the line-of-sight, in order to ensure the optimal geometry to detect the deformation. For Andromeda \citep{doliva23} and Centaurus~A \citep{muller18}, we have a good idea about the three-dimensional structure of their satellites systems. Then, one should be careful about the bright satellites, since their centers are not in the deep-MOND regime and therefore there will be no MOND-specific effects. Then, the satellites should not be too far away from the host. In \fig{shape}, Fornax from our simulation is shown near its pericenter, when the deformation is maximum. When it is near the apocenter, any asymmetry is hardly noticeable. Quantifying all these effects requires a dedicated study.

\begin{table}[b!]
\caption{Parameters of the disky Fornax 
in the simulation of tidal stirring.}
\label{tab:stir}
\centering
\begin{tabular}{lll}
\hline\hline                   
Parameter  & Value  & Note  \\
\hline \hline 
Scale length & 0.75\,kpc & (1) \\
Scale height & 0.23\,kpc & (2)\\
Truncation radius & 60\,kpc \\ 
Toomre Q  & 2 & (3) \\
\hline 
\end{tabular}
\tablefoot{(1) Radial density profile is an exponential disk. (2) Vertical density profile is the $\sech^2$ profile with a constant scale height. (3) A fixed value of Toomre Q was used across the disk.} 
\end{table}

\begin{figure*}[t!]
  \includegraphics[width=17cm]{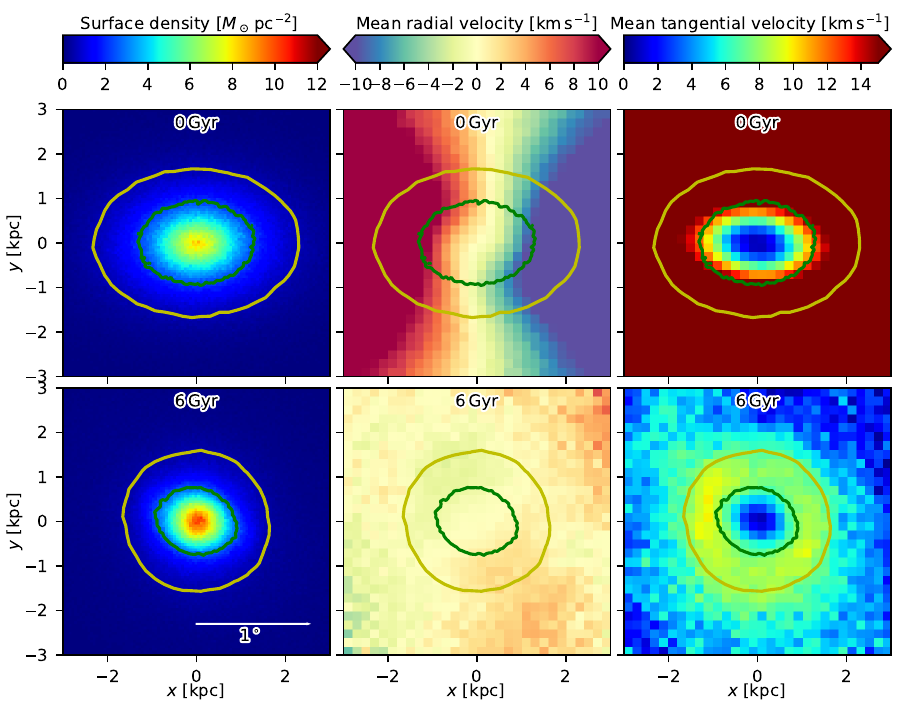}
     \caption{Transformation of rotating disky Fornax into a nonrotating spheroidal via tidal stirring. Top row: Beginning of the simulation. Bottom row: After 6\,Gyr. Left column: Surface density maps. The marked isophotes enclose 50\%\ and 80\% of the mass of the resulting object. Middle column: Maps of average line-of-sight velocity. Right column: Maps of the on-sky tangential velocity. All maps are views from the center of the MW. Fornax moves to the right. The bar in the lower left panel indicates one degree on the sky.}
     \label{fig:trans}
\end{figure*}

\section{Extra simulation: Transformation of a young disky Fornax into a nonrotating spheroid}
\label{sec:stir}
As argued in \sect{sanitytdg}, our scenario of the formation of Fornax implies that the galaxy had to transform from a rotating disk into a nonrotating spheroidal.  That disky dwarfs can transform into dwarf spheroidals as they orbit the MW has been demonstrated within the context of Newtonian gravity \citep{mayer01,mayer01b,lokas10}. The process is called tidal stirring. In this section,  we demonstrate by a dedicated simulation that such a transformation is possible for Fornax in MOND. We aim just for a rough reproduction of the properties of Fornax, since the main objective of this work is to investigate the survival of GCs in a spherical model of Fornax.

The simulation of the tidal stirring was the same as the simulations described before (\sect{sim}),  but Fornax was initially set as a rotating disk without GCs. The initial characteristics of the disky Fornax are listed in \tab{stir}. The disky model was generated using the MOND version  \citep{Banik_2020_M33} of the Disk Initial Condition Environment (\textsc{dice}; \citealp{Perret_2014}) code. After some experimentation, we found that tilting the spin of the disk in the direction $(1,0,-1)$ gives good results, which we describe below.

We investigated how Fornax changed after 6\,Gyr of orbiting the MW. The object lost about 30\% of its mass.  We quantified its shape by applying the singular value decomposition to particles which are within 3\,kpc of the Fornax barycenter to exclude particles in the stream. While Fornax was very flattened at the beginning of the simulation, having an axial ratio of 1:1:0.2, at 6\,Gyr it turned into a nearly spherical triaxial ellipsoid with an axial ratio of 1:0.9:0.8. 

Figure~\ref{fig:trans} shows how the simulated Fornax would be observed from the center of the MW at the beginning of the simulation and at 6\,Gyr. The change of shape is depicted in the first column, showing the maps of surface density. Marked are the contours enclosing 50 and 80\% of the mass of the final object. The final 50\% mass contour has a size that agrees well with the effective radius of the observed Fornax  of 0.75\,kpc. The same two contours are replicated in the other columns of the figure. The middle column shows the maps of the mean line-of-sight velocity. It shows that the initial strong rotation of the galaxy virtually disappeared. It is interesting that the remaining radial velocity pattern is not symmetric. This agrees with the complex kinematic structure of the real Fornax \citep{delpino17}. The right column of \fig{trans} shows that the initially very strong on-sky tangential velocity was largely eliminated too. The amplitude and appearance of the final tangential velocity field resemble the observed data \citep{martinezgarcia21}. As in the previous sections, we note that exact reproduction of the galaxy is not to be expected because our model ignores the fact that Fornax was initially gas rich, formed stars, and its density distribution was shaped by SN explosions.

\section{Conclusions}
\label{sec:sum}
Fornax possesses five classical GCs that are often used for testing dark matter and modified gravity theories. Here we use the GCs to test the QUMOND formulation of MOND using high-resolution $N$-body simulations in the popular PoR code.

\begin{itemize}
\item We found that to investigate whether the GCs of Fornax can survive, the wider context of the formation of Fornax has to be taken into account. The observations seem to imply that Fornax is an ancient TDG ejected during a MW-M\,31 encounter 7.5\,Gyr ago, and that its GCs started their lives as massive star clusters in the MW disk. This implies many useful pieces of information on how the simulations of the GC survival should be set up (\sect{obs}).

\item Our simulations show that, in QUMOND, the probability of the five GCs of Fornax surviving is 20\% (\sect{res}). This is cetainly not enough to exclude the theory.   
In agreement with previous analytic studies, we found that the GCs of Fornax would most probably sink in a few gigayears if they started at their observed projected positions \citep{ciotti04, sanchezsalcedo06, angus08}.

\item Fornax was initially gas rich and was forming stars until very recently. The SN explosions  could therefore have helped the GCs to survive through the mechanism described in \citet{bil24}. We found, however, that their effect is not sufficient or essential for explaining the observed positions of the GCs (\sect{res}). 

\item The simulations showed several other additional interesting facts. The orbit of Fornax around the MW predicted by MOND has pericenters that coincide with the bursts of star formation seen in the observed star formation history of Fornax (\sect{peaks}). This makes perfect theoretical sense, because tidal deformation is expected to trigger bursts of star formation. The simulations reproduce the observed velocity dispersion profile of Fornax (\sect{disp}), the spatial distribution of its GCs (\sect{gcdist}), and the existence of its stellar halo (\sect{halo}). The most massive GC of Fornax was able to provoke a feature in the galaxy similar to the observed so-called ``shell'' (\sect{shell}). 

\item The simulations also predict that there should be a faint stream extending from Fornax (\sect{stream}). Its observability would be at the limits of the current methods. The simulations also predict that faint satellites of giant galaxies should show MOND-specific shape deformations in the direction toward the host galaxy (\sect{lopsided}). While the geometry precludes the deformations from being observable for the MW satellites, looking for the effect among the satellites of M\,31, Centaurus~A, and other nearby giant galaxies appears more promising.

\item In our scenario, Fornax is  expected to  have started its life as a rotating disk galaxy, while it is observed to be a nonrotating spheroidal. Through an additional simulations, we demonstrated that the tidal influence of the MW is able to ensure the transformation (\sect{stir}).

\item One should be careful when using PoR. We found that in some settings, the code does not conserve the total angular momentum (\app{angmom}). For us, when the GC was simulated as a resolved $N$-body object, this caused an overly fast sinking of the GC.  

\end{itemize}

Despite these findings, it is important to acknowledge the limitations of this work. It would be desirable to revisit the problem with hydrodynamical simulations where the GCs initially orbit a rotating gas-rich disky Fornax. Ideally, the simulations should also include resolved GCs and the formation of Fornax out of a tidal arm.

\begin{acknowledgements}
This research was supported by the Ministry of Science, Technological Development and Innovation of the Republic of Serbia under contract no. 451-03-136/2025-03/200002 with the Astronomical Observatory of Belgrade.  HZ acknowledges support from the UK Science and Technology Facilities  Council grant ST/V000861/1, the USTC Fellowship for International Cooperation and the 111 Project for "Observational and Theoretical Research on Dark Matter and Dark Energy" (B23042).

\end{acknowledgements}

\bibliographystyle{aa}
\bibliography{literature}

\begin{appendix}

\section{Problems with conservation of angular momentum in PoR}
\label{app:angmom}

Initially, we modeled the GCs as resolved $N$-body objects instead of point masses. In such simulations, they experienced very fast sinking. Eventually, we found that this was because the total angular momentum of all particles in the box is not conserved for the simulations, which indicates a numerical problem. Since this problem has not been described in the literature on PoR yet, we present here more details as a warning. 

\begin{figure*}
  \includegraphics[width=17cm]{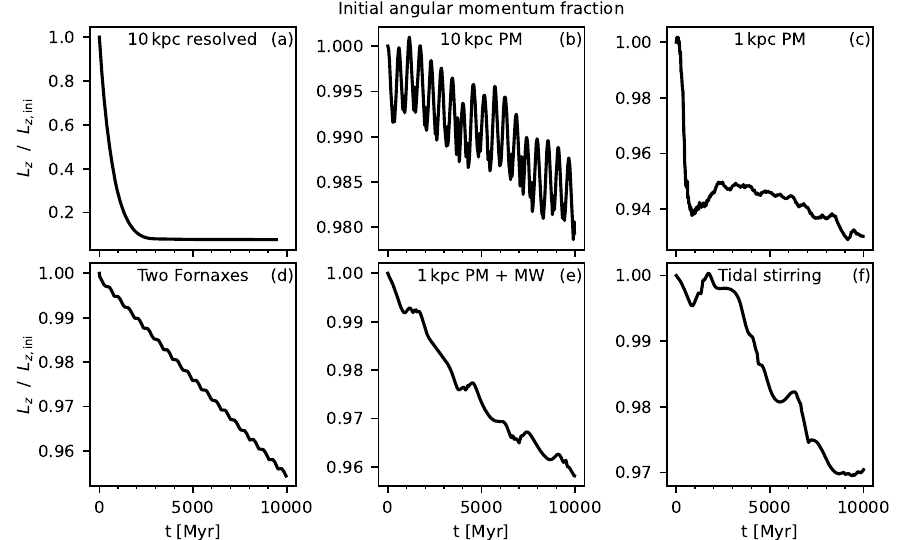}
     \caption{Test of conservation of angular momentum in different simulations. (a) Resolved GC orbiting initially on a circular orbit with a radius of 10\,kpc. (b) The same for point-mass GC. (c) Point mass GC orbiting initially on a circular orbit with a radius of 1\,kpc. (d) Two models of Fornax initailly orbiting each other on circular trajectories with a relative separation of 10\,kpc. (e) As in panel (c), but in addition the Fornax with the GC orbit a static MW. (f) Disky Fornax orbits a static MW in the simulation of tidal stirring (\sect{stir}). \vspace{4ex} }
     \label{fig:angmom}
\end{figure*}

The resolved GCs were modeled as Plummer spheres, with a mass of $10^5\,M_\sun$ and a scale radius of 3\,pc. They consisted of $2.5\times 10^4$ particles, that is each particle had $4\,M_\sun$ or 1/5 of the mass of the particles of Fornax (\sect{sim}). The model of the GC was prepared in the same way as the galaxy (\sect{sim}). The numerical setup was the one described in \tab{setup}, with the exception of setting {\sc m\_refine} = 20*80 and commenting out of the parameter {\sc mass\_sph}. In this way, the computational grid was refined if the cell contained at least 80 particles, regardless of their mass. 

Here we describe only a simulation where the resolved GC intially orbits on a circular trajectory around the $z$-axis with a radius of 10\,kpc, far beyond the last particle of Fornax (ca. 5\,kpc), and there is no MW. When the simulation is repeated with a point-mass GC, there is virtually no sinking within 10\,Gyr. But with a resolved GC, the sinking was virtually completed after only 2\,Gyr. The fact itself that the GC moves outside of the galaxy does not exclude dynamical friction to occur, as demonstrated by \citet{weinberg86}. Nevertheless, the true answer for why the resolved GC was sinking so quickly turned out to be the nonconservation of angular momentum in the simulation, as we demonstrate below.

We developed the following procedure to monitor for the conservation of the angular momentum in a simulation. For each output of the simulation, we first calculated the total angular momentum with respect to the corner of the computational box, $\vec{L}_\mathrm{raw}$. This quantity itself is difficult to interpret because the whole system always shows some numerical drift with respect to the computational grid. This numerical drift seems to be an accelerated motion with a lot of random noise. To remove this effect from the evaluation of the conservation of the total \textit{internal} angular momentum, we also calculated the angular momentum of the center of mass of the system with respect to the corner of the computational box, $\vec{L}_\mathrm{com}$. The total internal angular momentum that we are interested in is then $\vec{L} = \vec{L}_\mathrm{raw}-\vec{L}_\mathrm{com}$. We then evaluated the time dependence of the $z$-component of $\vec{L}$, that is $L_z$, as a function of time. For an easier interpretation,  $L_z$ was expressed in the units of the value of $L_z$ at the beginning of the simulation, denoted by $L_{z,\mathrm{ini}}$.

For the simulation with the resolved GC, all internal angular momentum is virtually lost after 2\,Gyr, as shown in panel (a) of \fig{angmom}. Even if an analogous simulation is run with Newtonian gravity in PoR, the situation is no better. The GC experiences virtually a free fall after a few tens of megayears.  On the other hand (speaking about the QUMOND gravity hereafter again), if the GC is modeled by a point mass, the simulation loses just 2\% of the initial angular momentum within 10\,Gyr (\fig{angmom}b). Similarly, if the point-mass GC is put on a circular orbit with a more relevant radius of 1\,kpc, the conservation of the total internal angular momentum is still relatively good: only about 6\% is lost during 10\,Gyr (\fig{angmom}c).

We also were interested in whether the problems with conservation of angular momentum pertain to simulations of interacting galaxies. We thus arranged two models of Fornax moving on a circular trajectory with an initial separation of 10\,kpc. All parameters of the code remained the same as in \tab{setup}. The simulation lost about 5\% of the initial total internal angular momentum within 10\,Gyr (\fig{angmom}d).  We speculate that the problem with a resolved GC stems from the very different timescales for the motion of the stars within the GC, within the galaxy, and the orbital time of the GC around the galaxy. 

The most important configuration to check for us are the simulations similar to those described in \sect{sim}. In those simulations, the particles of the MW are disabled to move with respect to the computational grid by using the {\sc static\_parts} version of PoR. This is justified because Fornax and its GCs have much smaller masses than the MW, and move much further away from the MW than the characteristic radius of the MW. To test the conservation of angular momentum, we used a simulation as from \sect{sim} where Fornax orbited the MW the GC initially moved around the $z$-axis on a circular prograde orbit around Fornax. The simulations indeed show a relatively good conservation of the total internal angular momentum, just about 4\% is lost (\fig{angmom}e).  In the simulation of tidal stirring (\sect{stir}), the loss is  3\% (\fig{angmom}f).

In total, we conclude that it is always better to check the conservation of angular momentum when using PoR, otherwise the simulation can produce a vastly incorrect result. Here we circumvented the problem by modeling  the GCs by point masses. It is possible that the problem could be solved by setting {\sc nexpand=2} and {\sc  n\_subcycle=1} in the namelist of PoR (suggestion by Romain Teyssier, the author of RAMSES, {\it priv. comm.}). The version of PoR we used however did not allow for setting these parameters.

\section{Influence of the other GCs}
\label{app:multigc}
\begin{figure}
        \resizebox{\hsize}{!}{\includegraphics{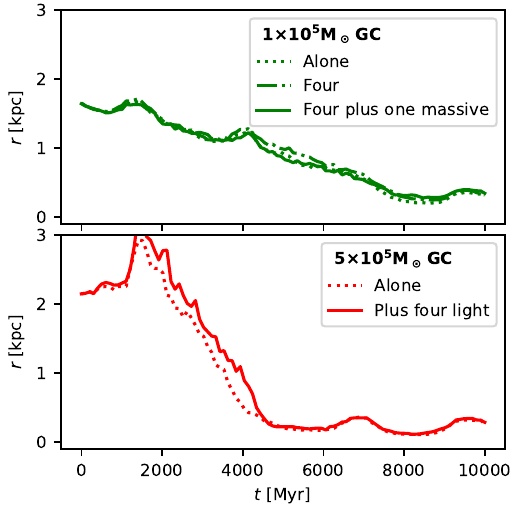}}
        \caption{Demonstration that the sinking time of GCs is, on average, not influenced much by the presence of the other GCs. 
        } 
        \label{fig:interactions}
\end{figure}

Our main study was based on simulations in which a there is only a single GC in Fornax. However, the real Fornax has five GCs. In this section, we find that simulating all five GCs at the same time has no substantial effect on the net sinking rate of the GCs. 

As in the rest of the paper, we assume that there is one massive GC in Fornax with the mass of $5\times10^5\,M_\sun$, and four light GCs, whose mass is $1\times10^5\,M_\sun$. First we made five reference simulations, each of which contained four light GCs with random initial conditions. The GCs started at random positions in a spherical layer between one and two kiloparsecs, each on a randomly oriented circular orbit. From \tab{results} we know that such GCs mostly survive or sink if are alone in Fornax. Then we made another series of simulations, such that one massive GC was inserted in each of the reference simulations. The goal was to see if the four light GCs influence the sinking the fifth massive one and vice versa, as the massive one sinks through the layer of the light GCs. The massive massive GC was lunched again on a randomly oriented circular orbit from between 1.5 and 2.5\,kpc, because when simulated isolated, such GC usually does not escape. 
Finally we made another set of simulations, where each of the GCs from the previous simulations was simulated being alone in Fornax with the same initial conditions. In each simulation, for each GC, we extracted for every timestep the between the GC and the center of Fornax.

To analyze the results, we sorted the different trajectories in these groups: 1) Massive GCs simulated alone; 2) light GCs simulated alone, 3) four light GCs simulated together (i.e., from the reference set), 4) the massive GCs in the simulations with five GCs, 5) the light GCs in the simulations with five GCs. For each group, we calculated for every time the median of distances of the GCs from the center of Fornax within the time window of $\pm250$\,Myr. The results are shown in \fig{interactions}. It shows that the presence of the other GCs does not have a substantial net impact on the sinking of the GCs. 

Interestingly, when the trajectories of the individual GCs are inspected, one can notice that the trajectory of a GC is influenced by the other GCs. This reflects the chaotic nature of the many-body problem: a small perturbation by the other GCs can grow substantially over time. Nevertheless, as demonstrated above, when averaged over many trajectories, the effect is very small.

\begin{table}
\caption{Influence of the uncertainty of the orbit of Fornax.}
\label{tab:unc}
\centering
\begin{tabular}{ll}
\hline\hline                   
Parameter  & Value               \\
\hline 
Pericenter [kpc] & $72 \pm 6$\\
Apocenter [kpc] & $153 \pm 3$\\
$g_\mathrm{ext, peri}~/~a_0$ & $0.13 \pm 0.01$\\
$g_\mathrm{ext, apo}~/~a_0$ & $0.058 \pm 0.001$\\
Min tidal radius [kpc] & $2.1 \pm 0.2$\\
Max tidal radius [kpc] & $4.54 \pm 0.08$\\
Last pericenter [Gyr ago] & $1.77 \pm 0.08$\\
$2^\mathrm{nd}$ last pericenter [Gyr ago] & $4.6 \pm 0.2$\\
$3^\mathrm{rd}$ last pericenter [Gyr ago] & $7.4 \pm 0.3$\\
$3^\mathrm{rd}$ last apocenter [Gyr ago] & $6.0 \pm 0.2$\\
\hline                                            
\end{tabular}
\end{table}

\setcounter{figure}{0}
\renewcommand{\thefigure}{D.\arabic{figure}}
\begin{figure*}[]
\centering
   \includegraphics[width=17cm]{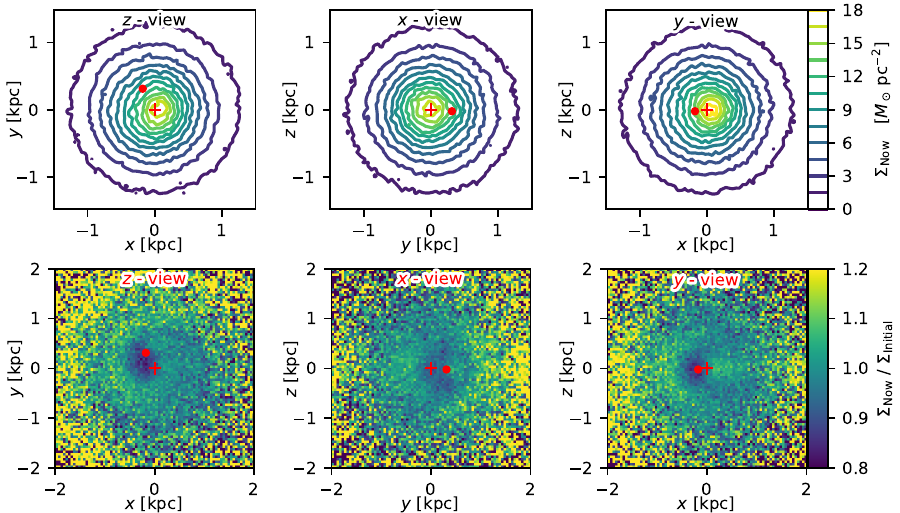}
     \caption{The ``shell'' in the simulated Fornax (\sect{shell}) seen along different axes of the simulation.}
     \label{fig:shellall}
\end{figure*}

\section{Uncertainty of the orbit of Fornax}
\label{app:unc}
In this section, we evaluate how the uncertainty  of the measurement of the position and velocity of Fornax relative to the MW propagates into the uncertainty of its orbit. We consider here the observational uncertainty of the heliocentric distance, radial velocity, and proper motion of Fornax, just as of the position and velocity of the Galactic center with respect to the Sun. We assumed that each of these quantities has a Gaussian distributions with the standard deviations given in the references stated in \tab{obs}. We integrated analytically backward the relative orbit of Fornax and the MW like in \sect{orbit} for 1000 random realizations of the initial conditions. For each orbit, we measured the quantities listed in \tab{unc}, which are those discussed in \sect{orbit}. For each quantity, this led to 1000 samples, from which we determined the 15.9, 50 and 84.1 percentiles to estimate the expected values and the one-sigma measurement uncertainties. The results are stated in \tab{unc}.

\section{Additional figures}
\label{app:add}

In this appendix, we present additional images of shell-like structure discussed in \sect{shell}. In \fig{shellall}, the object is displayed from different lines of sight.

\end{appendix}
\label{LastPage} 
\end{document}